\documentclass[%
 reprint,
superscriptaddress,
 amsmath,amssymb,
 aps,
 prx
]{revtex4-2}

\usepackage{graphicx}
\usepackage{dcolumn}
\usepackage{bm}
\usepackage{lipsum}
\usepackage{verbatim}
\usepackage{upgreek}

\usepackage{hyperref}

\def\DDDD{$D_\mathrm{3d}$}
\def\CHH{$C_\mathrm{2h}$}

\def\m{\textsuperscript{--}}
\def\ket#1{\left|{#1}\right\rangle}

\def\ud{\uparrow/\downarrow}

\def\doubleline#1#2{ \begin{tabular} [c]{@{}c@{}} {#1}\\ {#2} \end{tabular} }

\begin{document}

\preprint{APS/123-QED}

\title{Hyperfine Spectroscopy of Isotopically Engineered \\ Group-IV Color Centers in Diamond}

\author{Isaac B. W. Harris} 
\thanks{These two authors contributed equally}
\affiliation{Department of Electrical Engineering and Computer Science, Massachusetts Institute of Technology, Cambridge, Massachusetts 02139, USA}

\author{Cathryn P. Michaels}
\thanks{These two authors contributed equally}
\affiliation{Cavendish Laboratory, University of Cambridge, Cambridge CB3 0HE, United Kingdom}

\author{Kevin C. Chen}
\affiliation{Department of Electrical Engineering and Computer Science, Massachusetts Institute of Technology, Cambridge, Massachusetts 02139, USA}

\author{Ryan A. Parker}
\affiliation{Cavendish Laboratory, University of Cambridge, Cambridge CB3 0HE, United Kingdom}

\author{Michael Titze}
\affiliation{Sandia National Laboratories, Albuquerque, New Mexico 87123, USA}

\author{Jesús Arjona Martínez}
\affiliation{Cavendish Laboratory, University of Cambridge, Cambridge CB3 0HE, United Kingdom}

\author{Madison Sutula} 
\author{Ian R. Christen} 
\affiliation{Department of Electrical Engineering and Computer Science, Massachusetts Institute of Technology, Cambridge, Massachusetts 02139, USA}

\author{Alexander M. Stramma} 
\author{William Roth}
\author{Carola M. Purser}
\author{Martin Hayhurst Appel} 
\affiliation{Cavendish Laboratory, University of Cambridge, Cambridge CB3 0HE, United Kingdom}


\author{Chao Li} 
\affiliation{Department of Electrical Engineering and Computer Science, Massachusetts Institute of Technology, Cambridge, Massachusetts 02139, USA}

\author{Matthew E. Trusheim} 
\affiliation{DEVCOM, Army Research Laboratory, Adelphi, MD, 20783, USA}
\affiliation{Department of Electrical Engineering and Computer Science, Massachusetts Institute of Technology, Cambridge, Massachusetts 02139, USA}

\author{Nicola L. Palmer}
\author{Matthew L. Markham}
\affiliation{Element Six, Global Innovation Centre, Fermi Avenue, Harwell, Didcot, OX11 0QR, UK}

\author{Edward S. Bielejec}
\affiliation{Sandia National Laboratories, Albuquerque, New Mexico 87123, USA}

\author{Mete Atat\"{u}re} 
\email{ma424@cam.ac.uk}
\affiliation{Cavendish Laboratory, University of Cambridge, Cambridge CB3 0HE, United Kingdom}

\author{Dirk Englund} 
\email{englund@mit.edu}
\affiliation{Department of Electrical Engineering and Computer Science, Massachusetts Institute of Technology, Cambridge, Massachusetts 02139, USA}

\date{\today}

\begin{abstract}
A quantum register coupled to a spin-photon interface is a key component in quantum communication and information processing.
Group-IV color centers in diamond (SiV\m, GeV\m, and SnV\m) are promising candidates for this application, comprising an electronic spin with optical transitions coupled to a nuclear spin as the quantum register.
However, the creation of a quantum register for these color centers with deterministic and strong coupling  to the spin-photon interface remains challenging.
Here, we make first-principles predictions of the hyperfine parameters of the group-IV color centers, which we verify experimentally with a comprehensive comparison between the spectra of spin active and spin neutral intrinsic dopant nuclei in single GeV\m\ and SnV\m\ emitters.
In line with the theoretical predictions, detailed spectroscopy on large sample sizes reveals that hyperfine coupling causes a splitting of the optical transition of SnV\m\ an order of magnitude larger than the optical linewidth and provides a magnetic-field insensitive transition.
This strong coupling provides access to a new regime for quantum registers in diamond color centers, opening avenues for novel spin-photon entanglement and quantum sensing schemes for these well-studied emitters.

\end{abstract}
\maketitle

\section{\label{sec:intro}Introduction}

Spin-photon interfaces are the backbone for many quantum communication~\cite{Bernien2013, Kalb2017a, Pompili2021, Stas2022}, transduction~\cite{Neuman2021,Raniwala2022}, sensing~\cite{Staudacher2013a, Kim2019, Eisenach2021}, and computing schemes~\cite{Hucul2015, Choi2019}, with one of the most promising physical implementations being color centers in the solid-state~\cite{Atature2018MaterialTechnologies}.
Group-IV color centers in diamond are particularly appealing for spin-photon interfaces, offering excellent optical properties when integrated into nanostructures~\cite{Nguyen2019a, Machielse2019, Wan2020, Rugar2020Narrow-LinewidthWaveguide,ArjonaMartinez2022PhotonicDiamond}, insensitivity to charge noise~\cite{DeSantis2021}, and high-fidelity spin control~\cite{Pingault2017, Debroux2021}.
In these systems, localized electronic states around a lattice defect provide a local spin qubit, while the transitions between ground and excited electronic orbital energy levels provide optical access for readout, initialisation, and spin-photon entanglement generation~\cite{Becker2018All-OpticalTemperatures,Debroux2021,Nguyen2019}.

Central to many color-center-based quantum protocols is the presence of a local quantum register coupled strongly to the spin-photon interface, which allows for the storage of quantum information while spin-photon entanglement operations are being performed ~\cite{Michaels2021MultidimensionalRegister,Russo2019GenerationEmitters,Russo2018PhotonicCommunications}.
The most common implementation of this local register is with a proximal spin-active nucleus of the host lattice, coupled to the electron in the spin-photon interface by hyperfine coupling, such as $^{13}$C in diamond~\cite{Felton2009, Szasz2013, Defo2021}.
\textsuperscript{13}C memories have been used in demonstrations of quantum networks~\cite{Nguyen2019,Pompili2021,Abobeih2022,Bradley2022}, and even allow for tens of nuclear spin memories coupled to a single spin-photon interface~\cite{Bradley2019}. 
However, the inclusion of the \textsuperscript{13}C register is inherently non-deterministic, reducing the yield of good quantum registers.
Intrinsic dopant nuclear spins, on the other hand, allow for the deterministic inclusion of a quantum register~\cite{Pfaff2012, Stas2022}.
For many applications where a large number of quantum registers is not required, these properties make dopant nuclei easier to integrate into quantum systems that require high-yield, high-fidelity control.

While the electronic spin properties of group-IV color centers have been investigated~\cite{Neu2013,Hepp2014a,Jahnke2015,Thiering2018a}, the development of a thorough understanding of the coupling to the intrinsic nuclear spin register is needed.
Each group-IV element has at least one nuclear isotope with spin $I>0$, but only the spin-1/2 \textsuperscript{29}Si in the SiV\m\ has been used as a memory qubit~\cite{Stas2022}, with a reported ground-state hyperfine coupling of approximately 70 MHz~\cite{Rogers2014,Pingault2017}.
Hyperfine coupling in other group-IV color centers remains understudied, with only a few recent reports of hyperfine coupling, including to the intrinsic nuclear spin of GeV\m~\cite{Adambukulam2023}, to weakly coupled~\cite{Trusheim2018} and strongly coupled~\cite{Debroux2021} \textsuperscript{13}C nuclei using the SnV\m~ and to the intrinsic nuclear spin of \textsuperscript{117}SnV\m\cite{Parker2023}.

\begin{figure*}
    \centering
    \includegraphics[width=\textwidth]{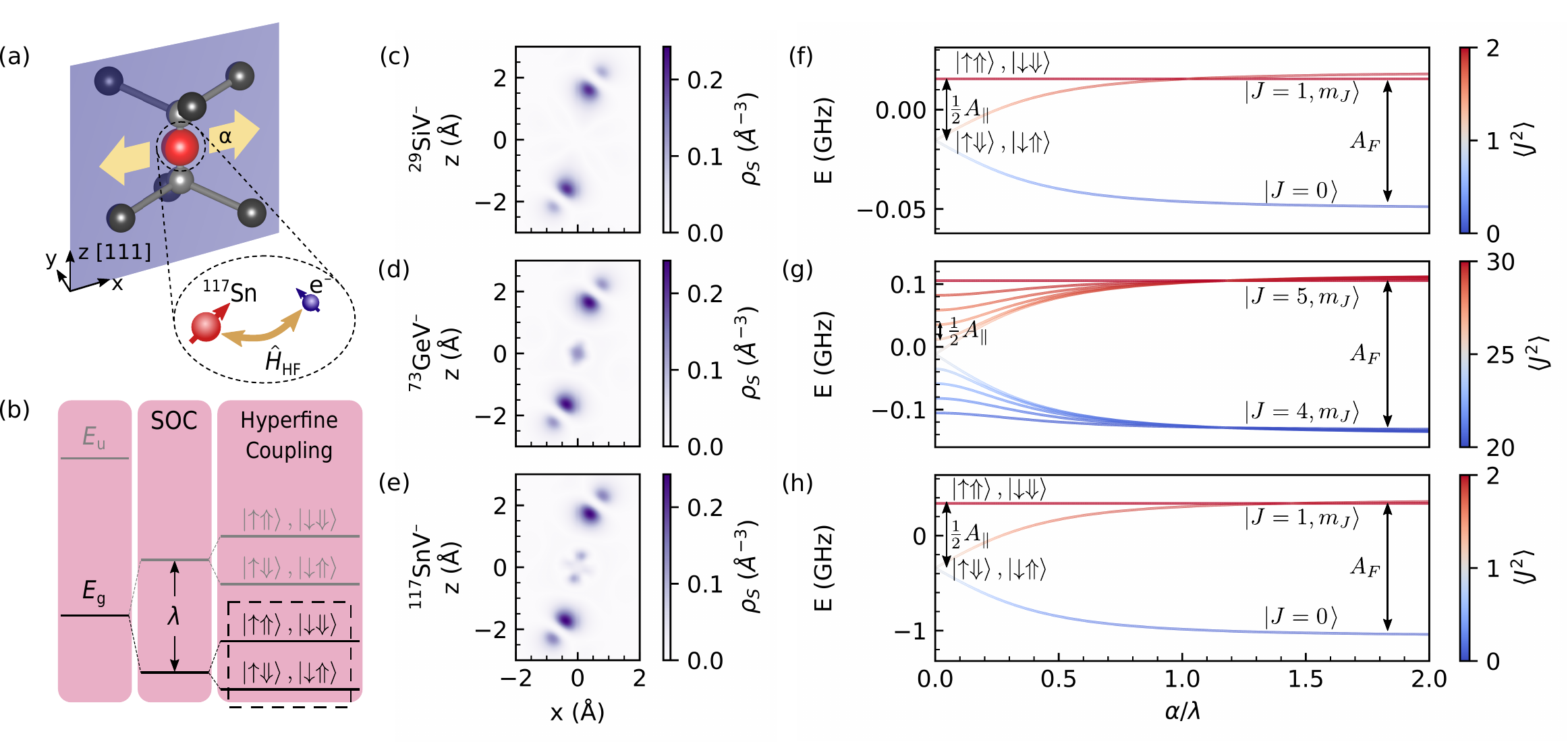}
    \caption{ First-principles hyperfine structure of group-IV color centers.
    (a) Split-vacancy structure of a representative group-IV color center, \textsuperscript{117}SnV\m, highlighting the interaction between the electron and nuclear spins.
    The group-IV dopant is shown in red, the nearest-neighbor carbon atoms in black and the lattice vacancies in gray.
    (b) Level structure showing $E_\mathrm{g}$ and $E_\mathrm{u}$ manifolds for a spin-1/2 group-IV color center.
    Spin-orbit coupling (SOC), $\lambda$, splits each manifold into two branches (only shown for $E_\mathrm{g}$).
    The addition of the hyperfine interaction with a spin-1/2 nucleus at zero strain splits each branch into two degenerate levels with aligned or anti-aligned nuclear and electron spins.
    (c, d, e) Cross-section of the spin density along the plane shown in panel a for \textsuperscript{29}SiV\m, \textsuperscript{73}GeV\m, and \textsuperscript{117}SnV\m.
    The presence of a heavier group-IV ion results in an increased spin density at the inversion symmetry center.
    (f, g, h) Energies of the lower branch ground state hyperfine levels as a function of strain, $\alpha$, shown in panel a and expressed as a fraction of spin-orbit coupling strength, for \textsuperscript{29}SiV\m, \textsuperscript{73}GeV\m, and \textsuperscript{117}SnV\m.
    Red and blue indicate the total angular momentum squared of the electro-nuclear system, $\langle J^2\rangle$, and correspond to the cases when electron and nucleus are aligned or anti-aligned respectively.
    }
    \label{fig:DFT}
\end{figure*}

In this paper, we develop a detailed model of the intrinsic nuclear memory of group-IV color centers and predict the hyperfine parameters for each group-IV isotope using density functional theory.
We then compare these predictions with our experimental observations of optical hyperfine signatures for \textsuperscript{73}GeV\m,  \textsuperscript{117}SnV\m, and \textsuperscript{119}SnV\m, allowing the assignment of isotopic signatures directly to the optical spectrum.
We demonstrate that SiV\m, GeV\m, and SnV\m\ have substantially higher hyperfine coupling strengths than NV\m, and we capture a proportionality between hyperfine strength with atomic mass of the color center.
This leads to an optical signature of SnV\m\ spin-active isotope hyperfine splitting that is 13 times larger than the homogeneous linewidth of the optical transitions, allowing for direct optical access of the nuclear spin.

\section{\label{sec:DFT}First-Principles Model of Group-IV Color Center Hyperfine Interaction}

Group-IV color centers consist of a single dopant atom from column 4 of the periodic table sitting at a \DDDD-symmetric interstitial site between two neighboring carbon vacancies in the diamond lattice, as shown in Fig.~\ref{fig:DFT}(a).
The electronic structure is a single hole orbiting the defect~\cite{Hepp2014a}, which exists in a ground $E_\mathrm{g}$ or excited $E_\mathrm{u}$ state, as illustrated in Fig.~\ref{fig:DFT}(b).
We model the state of the hole in either manifold as a tensor product of the orbital and spin degrees of freedom, resulting in a basis of the four states
\begin{equation}\label{eq:basis}
    \ket{ \psi_{\mathrm{gnd(exc)}} } = \ket{ e_\mathrm{g(u)\pm} } \otimes \ket{ \ud }
\end{equation}
where the quantisation axis of the orbital and spin degrees of freedom is along the \DDDD\ symmetry axis.
The Hamiltonian for the hole-spin system, $\hat{H}_\mathrm{S}$, is a sum of contributions from spin-orbit ($\hat{H}_\mathrm{SOC}$), strain ($\hat{H}_\mathrm{Egx/y}$), and magnetic-field ($\hat{H}_\mathrm{B/L}$) coupling.
Hence, 
\begin{equation}\label{eq:electron}
    \hat{H}_\mathrm{S} = \hat{H}_{\mathrm{SOC}} + \hat{H}_\mathrm{Egx} + \hat{H}_\mathrm{Egy} + \hat{H}_\mathrm{B}  + \hat{H}_\mathrm{L}
\end{equation}
where, in the basis defined in Eq.~\ref{eq:basis}, 
\mbox{$\hat{H}_{\mathrm{SOC}}=\frac{1}{2}\lambda\,\sigma_z^\mathrm{orb}\sigma_z^\mathrm{S}$},
\mbox{$\hat{H}_\mathrm{Egx/y}=-\alpha\sigma_{x/y}^\mathrm{orb}$},
\mbox{$\hat{H}_\mathrm{B}=g\mu_\mathrm{B} \boldsymbol{B}\cdot\hat{\mathbf{S}}$},
\mbox{$\hat{H}_\mathrm{L} = q\mu_\mathrm{B} B_z \sigma_z^\mathrm{orb}$},
$g$ is the electron g-factor, $\mu_\mathrm{B}$ is the Bohr magneton, \mbox{$\hat{\mathbf{S}}=\frac{1}{2}(\sigma_x^\mathrm{S}, \sigma_y^\mathrm{S}, \sigma_z^\mathrm{S})$} is the standard electron spin operator, and $\sigma_i^\mathrm{orb/S}$ are the Pauli matrices applied to the orbital/spin degree of freedom.
The forms of the four-level Hamiltonian for these perturbations are inferred from group theory~\cite{Tinkham1992, Doherty2011, Hepp2014a}, generally up to a constant factor that must be calculated from first-principles~\cite{Thiering2018a} or measured experimentally~\cite{Hepp2014a, Meesala2018a}.
We do not model separately the symmetry-breaking Jahn-Teller distortion, as its effects can be absorbed into an effective reduction of the spin-orbit interaction~\cite{Thiering2018a}.

\subsection{Hyperfine interaction model}
The nuclear spin interacts with magnetic field via the nuclear Zeeman interaction
\begin{equation}\label{eq:nucleus}
    \hat{H}_\mathrm{I}=g_\mathrm{I}\mu_\mathrm{N} \boldsymbol{B}\cdot\hat{\boldsymbol{I}}
\end{equation}
where $g_\mathrm{I}$ is the nuclear $g$-factor, $\mu_\mathrm{N}$ is the nuclear magneton, and $\hat{\boldsymbol{I}}$ is the standard nuclear spin operator for a spin $I$ nucleus.
To model the hyperfine interaction, we expand the basis in Eq.~\ref{eq:basis} to include the nuclear spin degree of freedom quantized along the defect axis of symmetry, $\ket{m_\mathrm{I}}$.
The nucleus interacts with the electronic spin-orbit system via the hyperfine interaction
\begin{equation}\label{eq:HF}
    \hat{H}_{\mathrm{HF}} = \hat{\mathbf{S}} \cdot \mathbf{A} \cdot \hat{\mathbf{I}}
\end{equation}
where $\mathbf{A} = \mathbf{A}_\mathrm{FC} + \mathbf{A}_\mathrm{DD}$ is the hyperfine tensor~\cite{Szasz2013, Defo2021}.
As discussed in Appendix~\ref{sec:dft_calculations}, the terms $\mathbf{A}_\mathrm{FC}$ and $\mathbf{A}_\mathrm{DD}$ correspond to the Fermi contact interaction and the dipole-dipole interaction respectively.

The Fermi contact is isotropic, and so can be expressed as a single scalar parameter $A_\mathrm{FC}$ multiplied by the identity matrix.
For a group-IV nucleus located at the inversion-symmetric point in the color center, the \DDDD\ symmetry restricts $\mathbf{A}_\mathrm{DD}$ to be a diagonal matrix with elements \mbox{$A_\mathrm{DD}=-2A_\mathrm{DD}^{xx}=-2A_\mathrm{DD}^{yy}=A_\mathrm{DD}^{zz}$} in the frame defined in Fig.~\ref{fig:DFT}(a)~\cite{Munzarova2004}.
Eq.~\ref{eq:HF} then becomes
\begin{equation}\label{eq:HF_reduced}
    \hat{H}_\mathrm{HF} = A_\perp \left( \hat{S}_x\hat{I}_x + \hat{S}_y\hat{I}_y \right) + A_\parallel\hat{S}_z\hat{I}_z.
\end{equation}
The $A_\parallel=A_\mathrm{FC}+A_\mathrm{DD}$ term is a shift of the energy levels that depends on the alignment of the electronic and nuclear spins.
The $A_\perp=A_\mathrm{FC}-2A_\mathrm{DD}$ terms mix states with identical orbital angular momentum but different spin and nuclear quantum numbers.
In the absence of other perturbations, the hyperfine levels form two hyperfine manifolds with total angular momentum $J=I\pm S$ separated in energy by $A_\parallel$, with further splitting $\frac{3}{2}A_\mathrm{DD}$ between the $m_\mathrm{J}$ sublevels within each manifold.

Nuclear spin-orbit and quadrupole coupling, discussed in Appendix~\ref{sec:hamiltonian_details}, also result in additional terms in the hyperfine interaction Hamiltonian.
These terms contribute less than 5\% of the total hyperfine interaction strength, so we do not include these in our model.
The Jahn-Teller distortion mentioned previously also affects the hyperfine interaction, however the change in interaction strength is less than 10\% of $A_\mathrm{FC}$ for SiV\m\ as discussed in Appendix~\ref{sec:hamiltonian_details}.
We also exclude this effect from our model on the grounds that it is not strong enough to cause qualitative change in the system.
This exclusion also allows us to avoid modeling the vibronic electron-photon modes for computational simplicity.

The final equation for the electro-nuclear system is the sum of the terms in Eq.~\ref{eq:electron}, Eq.~\ref{eq:nucleus}, and Eq.~\ref{eq:HF_reduced}:
\begin{equation}\label{eq:hamiltonian}
    \hat{H} = \hat{H}_\mathrm{S} + \hat{H}_\mathrm{I} + \hat{H}_\mathrm{HF}.
\end{equation}

\subsection{Hyperfine parameters from DFT}

To estimate the values of $A_\mathrm{FC/DD}$, we perform density functional theory (DFT) calculations with Quantum Espresso~\cite{QE1}, as detailed in Appendix~\ref{sec:dft_calculations}.
The resulting spin density in the ground state is plotted in Fig.~\ref{fig:DFT}(c-e) for SiV\m, GeV\m, and SnV\m\ respectively, and the calculated values for the hyperfine parameters are shown in Tab.~\ref{tab:summary}.

Normalizing for the nuclear gyromagnetic ratio, the hyperfine interaction increases with the mass of the group-IV dopant nucleus.
This is explained by the increasing contribution of the dopant orbitals compared to the carbon dangling bonds for the heavier elements.
The increased dopant orbital contribution to the spin density results in a large spin density near the dopant nucleus, and therefore an increased hyperfine interaction.
This general trend is also present in other systems, such as group-V donors in silicon~\cite{Feher1956, Pica2014}, and molecular systems~\cite{Knight1971}.

\subsection{Effect of strain on hyperfine structure}
The electronic spin-orbit interaction, $\hat{H}_\mathrm{SOC}$, contained within the $\hat{H}_\mathrm{S}$ term, separates states in which the hole-spin is aligned/anti-aligned with the orbital angular momentum by an amount $\lambda$, as shown in Fig.~\ref{fig:DFT}(b).
The hyperfine shift due to the $A_\parallel$ term remains unaffected by this spin-orbit splitting.
However, the splitting also means that the upper and lower branches contain states with opposite $e_\mathrm{g(u)\pm}$ orbital character.
As the $A_\perp$ terms can only mix states with the same orbital character, in the limit of $\lambda\gg A_\perp$ these terms are perturbatively suppressed.
The eigenstates cease to have well-defined total angular momentum, and are simply of the form $\ket{e_\mathrm{g(u)\pm}}\ket{\ud}\ket{m_\mathrm{I}}$, with eigenvalues separated by $\frac{1}{2}A_\parallel$, as can be seen in Fig.~\ref{fig:DFT}(b).

When transverse $E_\mathrm{gx(y)}$ strain is introduced into the system, as parameterized by the amount of strain $\alpha\,(\beta)$, the upper and lower orbital states are further mixed together.
The hyperfine levels are affected by the orbital mixing, as shown in Fig.~\ref{fig:DFT}(f-h) for $E_\mathrm{gx}$ strain using the DFT ground state values for \textsuperscript{29}SiV\m, \textsuperscript{73}GeV\m, and \textsuperscript{117}SnV\m.
$E_\mathrm{gy}$ strain only changes the orbital character of the orbitals in the upper/lower branch, and has an identical effect on the hyperfine energy levels.
In the limit of large strain $\alpha, \beta\gg\lambda$, the upper and lower branches have well-defined orbital character, and the $A_\perp$ hyperfine terms are again able to mix the states $\ket{\ud}\ket{m_\mathrm{I}}$ within each branch.
Each orbital branch then separates into two hyperfine manifolds of well-defined total spin angular momentum $J$, separated by $A_\mathrm{FC}$, and with a spacing of $\frac{3}{2}A_\mathrm{DD}$ between the $m_\mathrm{J}$ sublevels.
The isotopes \textsuperscript{29}Si and \textsuperscript{117}Sn are spin-1/2 nuclei, and as such they exhibit splitting into $J=0$ singlet, and $J=1$ triplet states, whereas the spin-9/2, \textsuperscript{73}Ge nucleus has a more complicated splitting into $J=4$ and $J=5$ states.

\subsection{Hyperfine structure in optical spectra}

The strength of the optical transitions is proportional to the matrix element $\langle\psi_\mathrm{exc}|\hat{d}|\psi_\mathrm{gnd}\rangle$, where $\hat{d}$ is the electric dipole operator.
\begin{figure}
    \centering
    \includegraphics[width=\columnwidth]{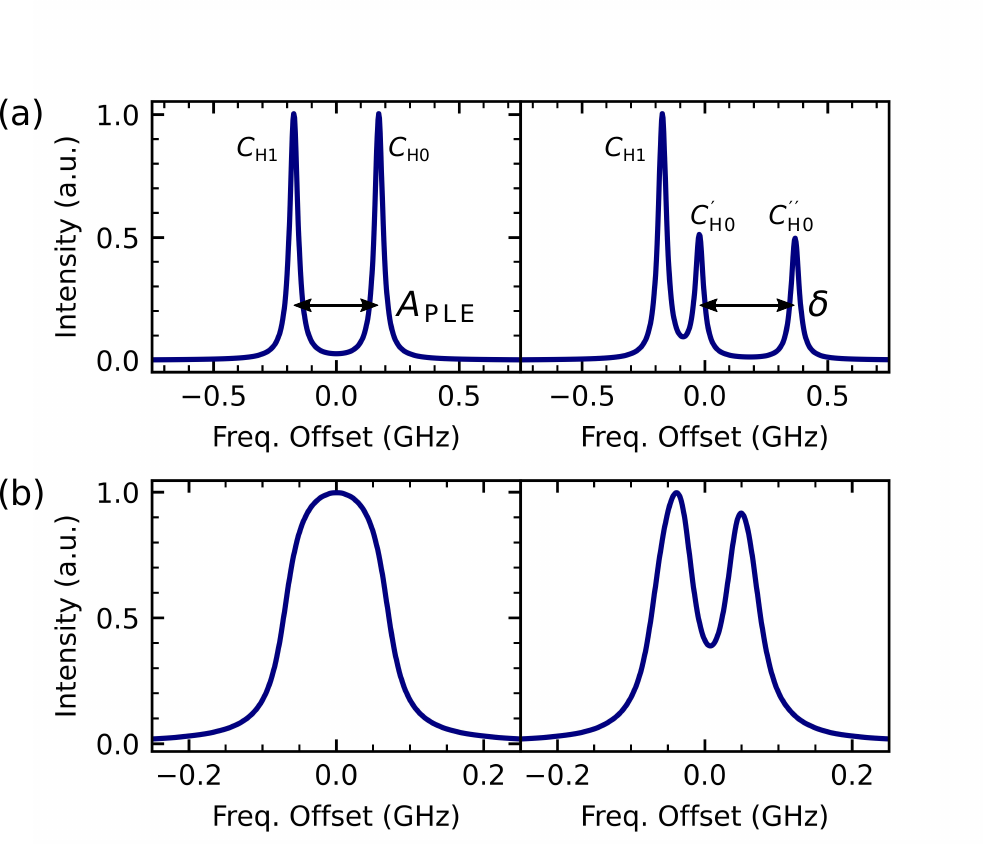}
    \caption{ Predicted spectra using DFT parameter and lifetime-limited linewidths for (a) \textsuperscript{117}SnV\m\ and (b) \textsuperscript{73}GeV\m\ at zero strain in the left panel, and a strain $\alpha/\lambda=0.15$ in the right panel.
    }
    \label{fig:spectra}
\end{figure}
Since $\hat{d}$ only acts on the orbital degree of freedom~\cite{Hepp2014a}, optical transitions cannot flip the nuclear or electronic spins, and only the spin-conserving transitions contribute significantly to the transition (see Appendix~\ref{sec:PLE_vis}).

We are in the limit of large spin-orbit coupling, where the electro-nuclear ground state Hamiltonian in Eq.~\ref{eq:hamiltonian} produces a series of equally spaced hyperfine levels in both the upper and lower branches of the ground state separated by $\frac{1}{2}A_{\parallel}^\mathrm{gnd}=\frac{1}{2}(A_\mathrm{FC}^\mathrm{gnd}+A_\mathrm{DD}^\mathrm{gnd})$.
Similarly, the excited state has hyperfine levels equally spaced by $\frac{1}{2}A_\parallel^\mathrm{exc}$.
The net result is that the electronic spin C-transition splits into a series of hyperfine transitions, with an optical hyperfine spacing \mbox{$A_\mathrm{PLE}=\frac{1}{2}(A_\parallel^\mathrm{exc} - \frac{1}{2}A_\parallel^\mathrm{gnd})$}.

For the spin-1/2 isotopes discussed in this paper, the hyperfine interaction results in four total transitions.
At zero strain, these occur in two degenerate pairs separated by $A_\mathrm{PLE}$: at lower frequency between the $m_\mathrm{J}=\pm 1$ hyperfine states in the ground/excited level ($C_\mathrm{H1}$), and between the two $m_\mathrm{J}=0$ states at a higher frequency ($C_\mathrm{H0}$).
The two peaks and their splitting $A_\mathrm{PLE}$ are labeled for \textsuperscript{117}SnV\m\ in Fig.~\ref{fig:spectra}(a).
The $m_\mathrm{J}=\pm1$ states are unaffected by strain, while the $m_\mathrm{J}=0$ states mix and separate in both ground and excited manifolds.
The strain-induced mixing splits the $C_\mathrm{H0}$ peak in the spectrum into two peaks, labeled $C_\mathrm{H0}'$ and $C_\mathrm{H0}''$, each having half the intensity of the $C_\mathrm{H1}$ peak, with a splitting $\delta$ labeled in Fig.~\ref{fig:spectra}(a).
The predicted hyperfine interaction for SnV\m\ is 10 times larger than the expected 35 MHz lifetime-limited linewidth of the transition~\cite{Trusheim2018}, making the hyperfine transitions directly resolvable.

Similarly for the spin-9/2 \textsuperscript{73}Ge isotope, we expect 20 hyperfine transitions, in 10 degenerate pairs which are equally spaced by $A_\mathrm{PLE}$.
The hyperfine parameters predicted by DFT for  \textsuperscript{73}GeV\m\ are small because of \textsuperscript{73}Ge's small gyromagnetic ratio compared to the other group-IV elements, with the Fermi contact parameter predicted to be 48~MHz in the ground state.
The resulting $A_\mathrm{PLE}=-13.78$ MHz is smaller than the expected lifetime-limited linewidth for GeV\m\ of 26 MHz~\cite{Bhaskar2017}, and the hyperfine level is therefore not predicted to be optically resolvable.
Nevertheless, the overlapping transitions result in a spectral line broadened by approximately $9|A_\mathrm{PLE}|=124$~MHz with a very non-Lorentzian flat-topped lineshape.
Strain splits the flat-topped emission peak into two peaks corresponding to the transitions between the $J=5$ levels at lower energy, and $J=4$ levels at higher energy, see Fig.~\ref{fig:spectra}(b).

The complete hyperfine model of the group-IV color centers discussed in this section gives a unique isotopic spectral signature for each emitter which we expect to be able to measure experimentally.
Some further discussion on the effects of strain on the hyperfine spectrum is found in Appendix~\ref{sec:PLE_vis}.

\section{\label{sec:PL}Isotope-Selective Spectroscopy}
\begin{table*}
    \centering
    \begin{tabular}{c|c|c|c|c|c|c|c|c|c}
        Isotope & Spin & $g_\mathrm{I}$ & \doubleline{$A_\mathrm{FC}^{gnd}$}{(DFT, MHz)} & \doubleline{$A_\mathrm{DD}^{gnd}$}{(DFT, MHz)} &
        \doubleline{$A_\mathrm{FC}^{exc}$}{(DFT, MHz)} &  \doubleline{$A_\mathrm{DD}^{exc}$}{(DFT, MHz)} &  \doubleline{$A_\mathrm{PLE}$}{(DFT, MHz)} & \doubleline{$A_\mathrm{PLE}$}{(exp., MHz)} & DFT error (\%)  \\ \hline
        \rule{0pt}{3ex}
        \textsuperscript{29}Si  & 1/2 & -1.110 &  64.20  &  -2.34 & -30.68 &  32.57 &  -29.98 &  -- & -- \\
        \textsuperscript{73}Ge  & 9/2 & -0.195 &  48.23  &  -1.35 &  5.03  &  14.30 &  -13.78 &  -12.5(5) & 9.2 \\
        \textsuperscript{115}Sn & 1/2 & -1.836 & 1275.04 & -24.47 & 386.74 & 230.43 & -316.70 &  -- & -- \\
        \textsuperscript{117}Sn & 1/2 & -2.000 & 1389.09 & -26.65 & 421.34 & 251.05 & -345.02 & -445(9) & 22 \\
        \textsuperscript{119}Sn & 1/2 & -2.092 & 1453.27 & -27.89 & 440.80 & 262.65 & -360.96 & -484(8) & 26 \\
    \end{tabular}

    \caption{Summary of hyperfine parameters predicted by DFT and measured experimentally}
    \label{tab:summary}
\end{table*}
To identify the spectroscopic signature of the group-IV color centers, we prepared diamonds plates with isotope-selective implantation of Si, Ge, and Sn, with isotope regions identified by the QR codes etched into the sample~\cite{Sutula2022} (see Appendix~\ref{sec:sample_prep} for more details).
A high-temperature annealing process combines the group-IV dopant with a vacancy to form the group-IV color center.

\begin{figure}
    \centering
    \includegraphics[width=\columnwidth]{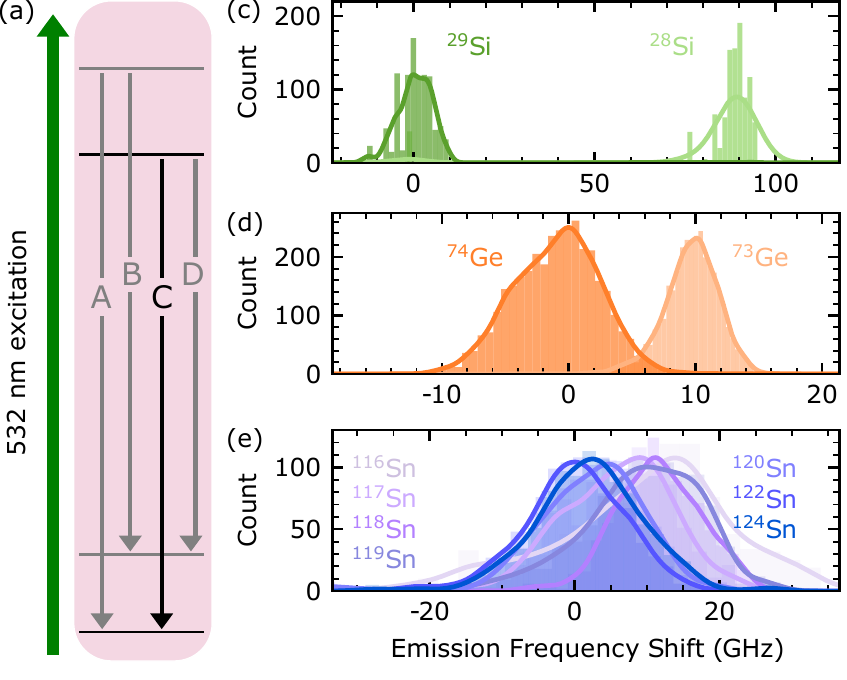}
    \caption{Characterization of isotope dependent photoluminescence.
    (a) Level structure of the group-IV color centers and photoluminescence transitions.
    (b-d) Distribution of the frequency shift of the C-transition for ensembles of SiV\m, GeV\m, and SnV\m. Solid lines are the Gaussian kernel density estimates. }
    \label{fig:Intro}
\end{figure}

Fig.~\ref{fig:Intro}(a), show the four possible transitions between the spin-orbit split levels of group-IV color centers, labeled A through D, which are optically addressable, with the exact frequency of these transitions depending on which group-IV dopant is present in the color center and the residual strain in the sample.
In addition, the exact isotope of the dopant also affects the transition energy, with a $\sqrt{M}$ dependence on isotope mass due to differences in the vibrational ground state energy~\cite{Dietrich2014IsotopicallyDiamond}.
To demonstrate this, we took photoluminescence (PL) measurements and fit a Gaussian lineshape to the C ZPL peak of several ensembles of emitters in order to extract the central frequency.

Fig.~\ref{fig:Intro}(b) shows the distribution of this extracted central PL frequency for both $^{28}$SiV\m and $^{29}$SiV\m. 
A clear shift of 83(8)~GHz in the central frequency can be seen, which is in good agreement with previous observation of 87~GHz~\cite{Dietrich2014IsotopicallyDiamond}.
This shift allows for the differentiation between the two isotopes without isotope selective implant and therefore the selection of a $^{29}$SiV center to make use of the spin-1/2 nucleus~\cite{Stas2022}. 

Fig.~\ref{fig:Intro}(c) shows the same measurement for $^{73}$GeV\m and $^{74}$GeV\m.
A shift of 13(7)~GHz can be seen, in line with previous experimental results of 15 GHz~\cite{Ekimov2015,Ekimov2017} and prediction based on the expected $\sqrt{M}$ dependency~\cite{Dietrich2014IsotopicallyDiamond}.
This means distinguishing single GeV centers using non-resonant excitation is unlikely, given the GeV centers typically have an inhomogeneous distribution of emission of several 10s of GHz~\cite{Iwasaki2015Germanium-VacancyDiamond}.

Similar measurements across 7 isotopes of SnV are shown in Fig.~\ref{fig:Intro}(d). The shift of the zero phonon line (ZPL) is hidden within the inhomogeneous distributions of each isotope.
The model and previous experimental results would suggest a shift between neighboring isotopes of $\approx$10~GHz~\cite{Narita2023}.
This is below the resolution of the spectrometer used to measure PL in this experiment, which combined with inhomogeneous distribution and non-resonant power broadening~\cite{Gorlitz2022CoherenceDiamond}, masks the isotope shift. The shift between isotopes further apart in mass is also hidden, likely due to the imperfect mass selectivity which brings these ensemble PL distributions towards a central value. 


\section{Hyperfine Photoluminescence Excitation Spectra}\label{sec:PLE}
To quantify the hyperfine parameters experimentally, we performed photoluminescence excitation (PLE) on the isotope-selectively implanted samples (see Appendix~\ref{sec:ple_methods}).
We collected statistics on color centers by performing wide-field PLE (WFPLE) experiments~\cite{Sutula2022} on regions of the sample with \textsuperscript{73}GeV\m, \textsuperscript{74}GeV\m, \textsuperscript{117}SnV\m, \textsuperscript{118}SnV\m, \textsuperscript{119}SnV\m and \textsuperscript{120}SnV\m\, with an implant dose of approximately 100 ions per site~\cite{Titze2022}.
This dose is sufficiently low that each site should only contain on the order of 1-5 emitters given color center creation yields of approximately 1-5\%~\cite{Wan2020}.
This may be further improved with recently developed in-situ photoluminescence spectroscopy for the formation of deterministic yield emitter arrays~\cite{Chandrasekaran2023}.
Given the large inhomogeneous linewidth compared to the typical homogeneous linewidth, we expect fewer than 0.1\% of emitters to have spectrally overlapping emission peaks within the same site. 

\subsection{\label{sec:GeV}PLE Spectroscopy of GeV}

\begin{figure}
    \centering
    \includegraphics[width=\columnwidth]{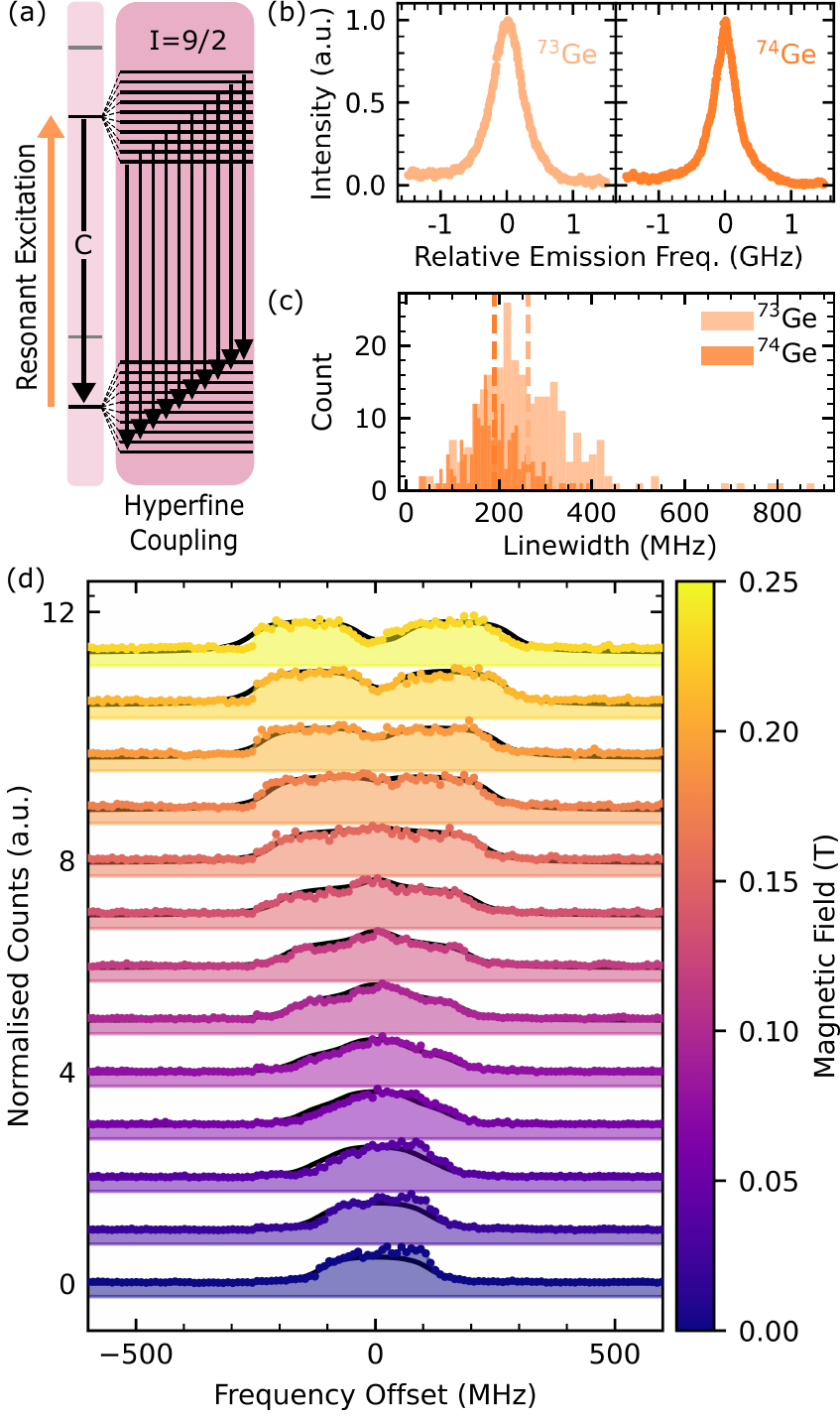}
    \caption{Hyperfine characterisation of GeV\m\ through PLE.
    (a) Hyperfine level structure of the spin-9/2 \textsuperscript{73}GeV\m\ electro-nuclear system.
    (b) Average WFPLE spectrum for \textsuperscript{73}GeV\m\ and \textsuperscript{74}GeV\m.
    (c) Distribution of emitter linewidths for \textsuperscript{73/74}GeV\m\ showing the larger linewidth of \textsuperscript{73}GeV\m\ .
    (d) PLE spectrum of a \textsuperscript{73}GeV\m\ emitter in the low power regime, as a function of magnetic field at 33$^\circ$ to the high symmetry axis, showing a tophat lineshape.
    Fit in black.}
    \label{fig:GeV}
\end{figure}

For the I=9/2 nucleus of \textsuperscript{73}GeV\m, we expect ten overlapping spin conserving transitions at zero field, as shown in Fig.~\ref{fig:GeV}(a) and discussed in Section~\ref{sec:DFT}.
When comparing the average WFPLE spectra of 242 \textsuperscript{73}GeV\m\ emitters and 195 \textsuperscript{74}GeV\m\ emitters in Fig.~\ref{fig:GeV}(b) we see that the \textsuperscript{73}GeV\m\ spectrum is substantially broader.
In Fig.~\ref{fig:GeV}(c), we plot a histogram of the Lorentzian linewidth fits of the \textsuperscript{73/74}GeV\m\ emitters, and see that the average linewidth for \textsuperscript{73}GeV\m\ is 262(7)~MHz, roughly 70 MHz broader than the 190(5)~MHz average linewidth for \textsuperscript{74}GeV\m\ under the same laser power.

To confirm that this broadening is a result of overlapping hyperfine transitions, we performed confocal PLE using chirped optical resonance excitation (CORE)~\cite{ArjonaMartinez2022PhotonicDiamond} (see Appendix~\ref{sec:ple_methods}) on a \textsuperscript{73}GeV\m emitter at varying magnetic field, shown in Fig. \ref{fig:GeV}(d).
We fit this data using the optical model of the isotope developed in Section~\ref{sec:DFT}, varying only the linewidth and $A_\mathrm{PLE}$.
These are found to be 72(3)~MHz and 12.5(5)~MHz respectively, within 10\% of the DFT prediction.
At zero field, the PLE spectrum with near-lifetime limited linewidth exhibits a flat-topped line shape due to the overlapping transitions (see Appendix~\ref{sec:PLE_vis}).
As the magnetic field is increased, the electro-nuclear levels separate into spin-aligned and spin-anti-aligned groups of energy levels.
At around 0.1 T only the transitions from the two groups near zero detuning still overlap, and a characteristic central hump surrounded by two broad shoulders appears, further highlighting that this lineshape comes from multiple overlapping hyperfine transitions.

\subsection{\label{sec:SnV}PLE Spectroscopy of SnV}

\begin{figure*}
    \centering
    \includegraphics[width=\textwidth]{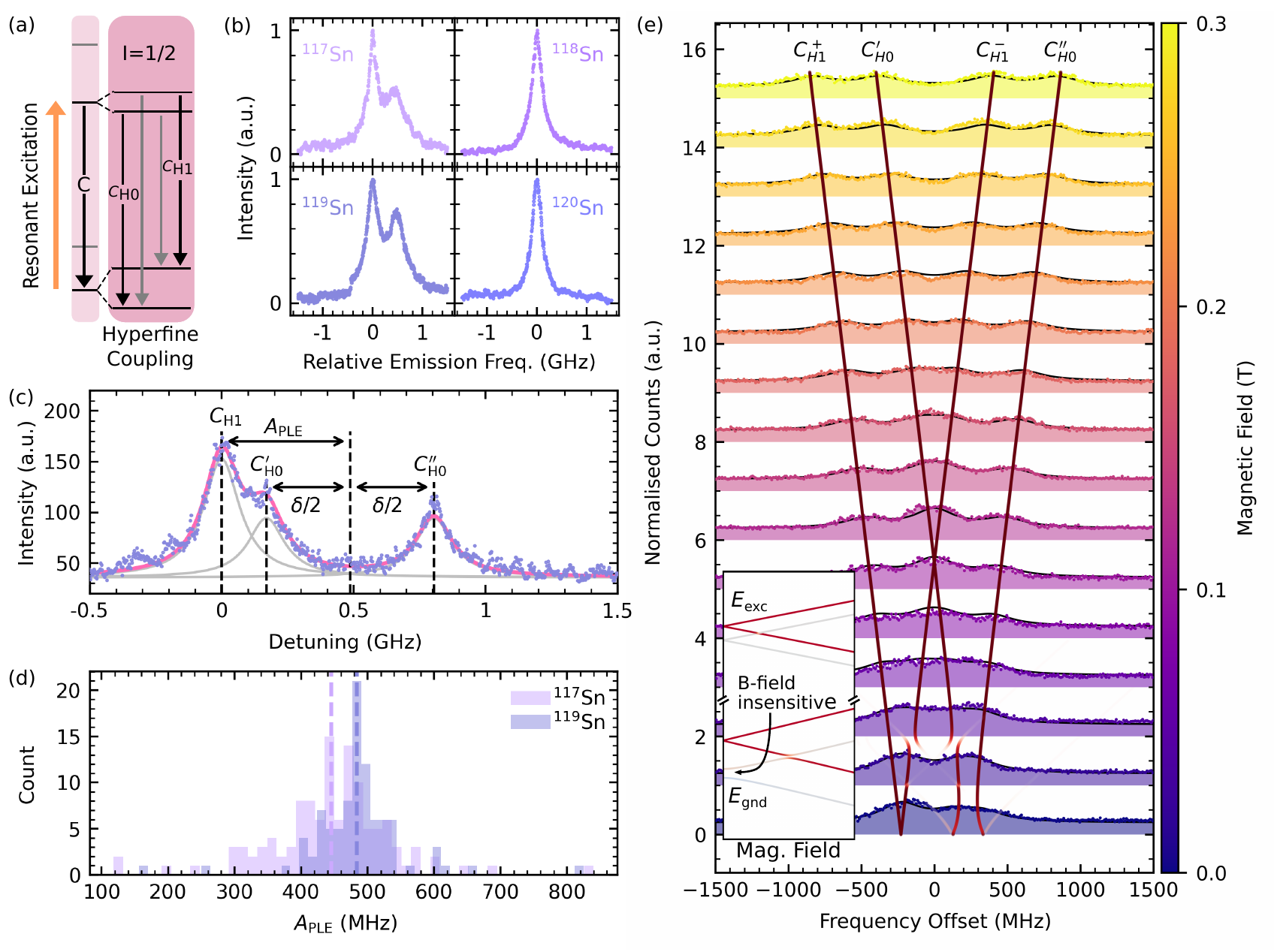}
    \caption{Hyperfine characterization of SnV\m\ via PLE.
    (a) Hyperfine level structure of a spin-1/2 SnV\m\ electro-nuclear system.
    (b) Averaged WFPLE spectrum for \textsuperscript{117}SnV\m, \textsuperscript{118}SnV\m, \textsuperscript{119}SnV\m, and \textsuperscript{120}SnV\m.
    (c) Typical WFPLE spectrum (purple) of a strained \textsuperscript{119}SnV\m, showing the hyperfine peaks with relative heights 2:1:1 (gray), split by the hyperfine parameter $A_\mathrm{PLE}$ and strain parameter $\delta$, with the combined fit with all transitions in pink.
    Peaks are labeled $C_\mathrm{H0}'$, $C_\mathrm{H0}''$  and $C_\mathrm{H1}$ as explained in the text.  
    (d) Histogram of the optical hyperfine splitting $A_\mathrm{PLE}$ for the spin-1/2 nuclei \textsuperscript{117/119}SnV\m.
    (e) PLE spectrum of a \textsuperscript{117}SnV\m\ emitter as a function of magnetic field, with fit in black, showing the three peaks at zero field split into four peaks labeled $C_\mathrm{H1}^+$, $C_\mathrm{H1}^-$, $C_\mathrm{H0}'$, and $C_\mathrm{H0}''$.
    Transition frequencies from the fit are overlaid in red, highlighting the avoided crossing of the $C_{H0}$ transitions near zero-field.
    Inset shows an illustration of the ground/excited level structure with avoided crossings at zero magnetic field.
    }
    \label{fig:SnV}
\end{figure*}

As discussed in Section~\ref{sec:DFT}, for SnV\m\ spin-1/2 isotopes we expect multiple spectrally distinct hyperfine transitions directly observable in the PLE spectra, as illustrated in Fig.~\ref{fig:SnV}(a).
This is shown experimentally in the average WFPLE spectra averaged over approximately 100 emitters for each of the isotopes \textsuperscript{117}SnV\m, \textsuperscript{118}SnV\m, \textsuperscript{119}SnV\m, and \textsuperscript{120}SnV\m\ in Fig.~\ref{fig:SnV}(b).
It is clear that additional spectral features appear for the two spin-1/2 isotopes \textsuperscript{117}Sn and \textsuperscript{119}Sn that are not present for the spin-0 isotopes \textsuperscript{118}Sn and \textsuperscript{120}Sn.

\begin{table}
    \centering
    \begin{tabular}{c|c|c|c|c}
        Isotope & Spin & \doubleline{Number of}{emitters.} &
        \doubleline{Fraction w/Hyper-}{fine Peaks (\%)}
          & \doubleline{$A_\mathrm{PLE}$ (S.E.)}{(MHz)} \\ \hline
          \rule{0pt}{3ex}
        \textbf{\textsuperscript{117}Sn} & \textbf{1/2} &\textbf{ 136 }& \textbf{87.5} & \textbf{445(9)} \\
        \textsuperscript{118}Sn &  0  & 119 & 16.0 & --\\
        \textbf{\textsuperscript{119}Sn} & \textbf{1/2} & \textbf{109} & \textbf{84.4} & \textbf{484(8)} \\
        \textsuperscript{120}Sn &  0  &  93 &  6.4 & --
    \end{tabular}

    \caption{Hyperfine statistics for the different SnV\m\ isotopes.}
    \label{tab:SnV}
\end{table}

We fit the PLE spectrum for each SnV\m\ individually with either a single Lorentzian peak (representing a spin-0 isotope), or three Lorentzian peaks with a 2:1:1 intensity ratio (corresponding to a spin-1/2 isotope).
For each spin-1/2 isotope fit, we extract the parameters $A_\mathrm{PLE}$ and $\delta$, as illustrated for a representative emitter in Fig.~\ref{fig:SnV}(c).
The 3-peak hyperfine feature is observed for more than 80\% of the emitters in the two spin-1/2 isotope-implanted regions, whereas it is present in less than 20\% of the emitters in the spin-0 isotope-implanted regions (see Tab.~\ref{tab:SnV}).
Performing a $\chi^2$ test on the number of emitters with the multi-peak PLE spectra in the spin-0 vs spin-1/2 regions, we conclude with a high degree of certainty that the multi-peak PLE is associated with the spin-1/2 isotopes ($p<10^{-5}$).
The bulk of the emitters of the wrong type likely come from imperfect isotope separation during the implantation (see Appendix~\ref{sec:sample_prep}).
We therefore assign the multi-peak feature to the spin-1/2 isotopes of tin.
This result is also in line with measurements performed on \textsuperscript{117}SnV\m\ in nanophotonic structures~\cite{Parker2023}, and clarifies a previous report of SnV\m\ hyperfine interaction strength of 40 MHz~\cite{Debroux2021}.
We now assess this later report as being due to the hyperfine coupling to a nearest-neighbor \textsuperscript{13}C nucleus, since it is closer in magnitude to previous predictions for \textsuperscript{13}C hyperfine coupling~\cite{Defo2021}.

The distribution of the hyperfine parameter $A_\mathrm{PLE}$ for \textsuperscript{117}SnV\m\ and \textsuperscript{119}SnV\m\ are shown in Fig.~\ref{fig:SnV}(d), and the mean value is summarized for all isotopes in Tab. \ref{tab:SnV}.
We note that $A_\mathrm{PLE}$ is larger for \textsuperscript{119}Sn than it is for \textsuperscript{117}Sn.
This is to be expected since the hyperfine coupling parameters are directly proportional to the nuclear gyromagnetic ratios~\cite{Szasz2013}.
We find the ratio \mbox{$A_\mathrm{PLE,^{119}Sn}/A_\mathrm{PLE,^{117}Sn}=1.09(0.04)$} to be in agreement with \mbox{$g_\mathrm{^{119}Sn}/g_\mathrm{^{117}Sn}=1.05$}.

We next performed PLE measurements at varying magnetic field on a strained \textsuperscript{117}SnV\m within a different device, detailed in Appendix~\ref{sec: dil fridge}, with results shown in Fig.~\ref{fig:SnV}(e).
We fit this data using the hyperfine model from Section~\ref{sec:DFT}, finding $A_\mathrm{PLE}=-459(3)$~MHz, $\alpha=55(3)$~GHz, and a power broadened linewidth of $336(3)$~MHz.
The simpler level structure compared to \textsuperscript{73}GeV\m\ allows us to directly track the trajectory of the transitions as a function of field strength.
The magnetic field breaks the degeneracy of the $C_{H1}$ peak's two constituent transitions, producing two peaks labeled $C_{H1}^\pm$, corresponding to transitions between the ground and excited $m_\mathrm{J}=\pm1$ total angular momentum states.
Due to the combined effect of an anti-crossing of the $m_\mathrm{J}=0$ ground states at zero magnetic field and the much weaker anti-crossing of the $m_\mathrm{J}=0$ excited states (see Fig.~\ref{fig:SnV}(e) inset), the $C_{H0}$ transitions exhibit an anti-crossing near zero-field.
For a sufficiently strained emitter, such as the \textsuperscript{119}SnV\m\ in Fig.~\ref{fig:SnV}(c), the strong coupling maintains optical access to the hyperfine levels at this ground level anti-crossing point.
Operating at this anti-crossing makes the levels magnetically insensitive to first order, suppressing the effect of magnetic noise~\cite{Bollinger1985, Kindem2020, Onizhuk2021}.
Nuclear spin bath magnetic noise has been shown to have a large effect on the coherence of group-IV color centers in previous work~\cite{Becker2018, Nguyen2019, Debroux2021}, so operating in such a regime may improve coherence.

\section{\label{sec:conc}Conclusions and discussion}
Our microscopic model of the electro-nuclear system and the first-principles calculations present a full theoretical framework for understanding the hyperfine coupling of the negatively charged group-IV color centers in diamond.
We predict a large increase in the hyperfine parameters moving down the group-IV column of the periodic table due to the increasing contribution of the dopant orbitals to the spin density.
We further show that spin-orbit coupling and strain must both be accounted for when modeling the resulting hyperfine levels.

Using isotope-selective ion implantation of group-IV elements in diamond, we are able to identify the optical hyperfine signatures of \textsuperscript{73}GeV\m, \textsuperscript{117}SnV\m, and \textsuperscript{119}SnV\m.
In particular, the spin-active SnV\m\ color centers show a clearly resolvable optical multi-peak feature compared to the spin-neutral isotopes due to the large hyperfine coupling to the intrinsic tin nucleus.
The hyperfine parameters and resulting PLE features predicted from DFT, as well as the measured PLE features are shown in Table \ref{tab:summary}.
The first-principles predictions are in sufficiently good agreement (within $\sim$20\%) to prove instructive in identifying the hyperfine signatures.
The remaining discrepancy between theory and experiment is likely due to the simplifications made in the modeling, such as the exclusion of Jahn-Teller distortion, and the use a local functional instead of a more sophisticated hybrid functional.

Both the GeV\m\ and SnV\m\ nuclear registers bring interesting challenges and opportunities for quantum information applications.
The spin-9/2 memory in \textsuperscript{73}GeV\m\ has a 10-level nuclear memory, allowing for the possibility of the generation of large cluster states by making use of the local memory~\cite{Michaels2021MultidimensionalRegister}.
Its nuclear quadrupole moment also means that the nuclear spin can potentially be driven directly by an electric field~\cite{Asaad2019}.

While the SnV\m\ isotopes have a more conventional spin-1/2 intrinsic memory, the strong hyperfine coupling means that the hyperfine levels can be accessed optically at zero field.
With non-zero strain applied to the defect at this zero-field operating point, our model predicts a magnetic-field insensitive transition between the $\ket{J=0/1,m_\mathrm{J}=0}$ ground states.
These states could thus combine the key attributes of a spin-photon interfaces: strong, direct, and stable optical transitions; environmentally insensitive hyperfine `clock states'; and additional hyperfine states for quantum memory.
The direct optical access of the hyperfine levels also allows for the possibility of direct transfer of photon states to the nuclear memory without using the electron spin as an intermediary, which may enhance spin-photon entanglement fidelity compared to existing schemes~\cite{Stas2022}.
Optical initialization and readout of the nuclear spin via this hyperfine optical transition has been demonstrated in a separate work~\cite{Parker2023}.

The presence of the strongly coupled memory in the well-established group-IV color center platform will allow future experiments to leverage their bright, high-quality optical emission in a new regime of quantum experiments.
The clear identification of the hyperfine parameters of GeV\m\ and SnV\m\ in this paper therefore sets the implementation roadmap for future work to use these nuclear spins as local memories for quantum information applications.

More generally, the theory and experimental methods developed here also present a new route to tailor other spin-photon interfaces where the interplay between spin-orbit coupling, crystal strain, and hyperfine interaction may also be important.
These include color centers in other materials systems, such as silicon~\cite{Prabhu2023, Higginbottom2022a}, silicon carbide~\cite{Miao2019, Wolfowicz2019, Crook2020}, rare-earth elements implanted in solids~\cite{Kindem2020}, and other emerging material platforms~\cite{Wang2023}.
This paper demonstrates that the selection of dopant isotopes can have a large effect on the resulting color center properties through the hyperfine structure.
Given the wide availability of isotopically-selective implantation, we see this as a useful tool and future standardized step in the fabrication of spin-photon interfaces.
\\
\\
\begin{acknowledgements}
This work was supported in part by the STC Center for Integrated Quantum Materials (CIQM) NSF Grant No. DMR-1231319, the National Science Foundation (NSF) Engineering Research Center for Quantum Networks (CQN) awarded under cooperative agreement number 1941583, and the MITRE Moonshot Program. 
We acknowledge support from the ERC Advanced Grant PEDESTAL (884745), the EU Quantum Flagship 2D-SIPC.
C.P.M. acknowledges support from the EPSRC DTP, R.A.P from the General Sir John Monash Foundation and a G-research Grant, J.A.M. from the Winton Programme and EPSRC DTP and A.M.S. from EPSRC/NQIT.
This work was performed, in part, at the Center for Integrated Nanotechnologies, an Office of Science User Facility operated for the U.S. Department of Energy (DOE) Office of Science. Sandia National Laboratories is a multimission laboratory managed and operated by National Technology \& Engineering Solutions of Sandia, LLC, a wholly owned subsidiary of Honeywell International, Inc., for the U.S. DOE’s National Nuclear Security Administration under contract DE-NA-0003525. The views expressed in the article do not necessarily represent the views of the U.S. DOE or the United States Government. 
M.S. acknowledges support from the NASA Space Technology Graduate Research Fellowship Program. 
We would additionally like to thank Hamza Raniwala and Hyeongrak Choi for helpful discussion.

I.B.W.H and C.P.M. contributed equally to this work.
\end{acknowledgements}

\bibliography{apssamp}

\newpage
\appendix

\section{Sample Preparation}\label{sec:sample_prep}
We prepared two samples UC1 and UC2 from single-crystal diamond plates ([N]$<5~$ppb) purchased from Element Six.
We removed the top 7~$\mu$m of diamond with Ar/Cl$_2$ and O$_2$ reactive ion etch (RIE) to relieve the strained surface layer.
Subsequently, we deposited 180~nm silicon nitride as a hardmask and patterned alignment markers and QR codes~\cite{Sutula2022} using electron beam lithography.
A thin layer of gold ($\sim 50~$nm in thickness) was then deposited and lifted off in HF acid, leaving metal covering only the alignment markers for optimal imaging contrast crucial for alignment during the subsequent ion implantation step.

\begin{figure*}
    \centering
    \includegraphics[width=\textwidth]{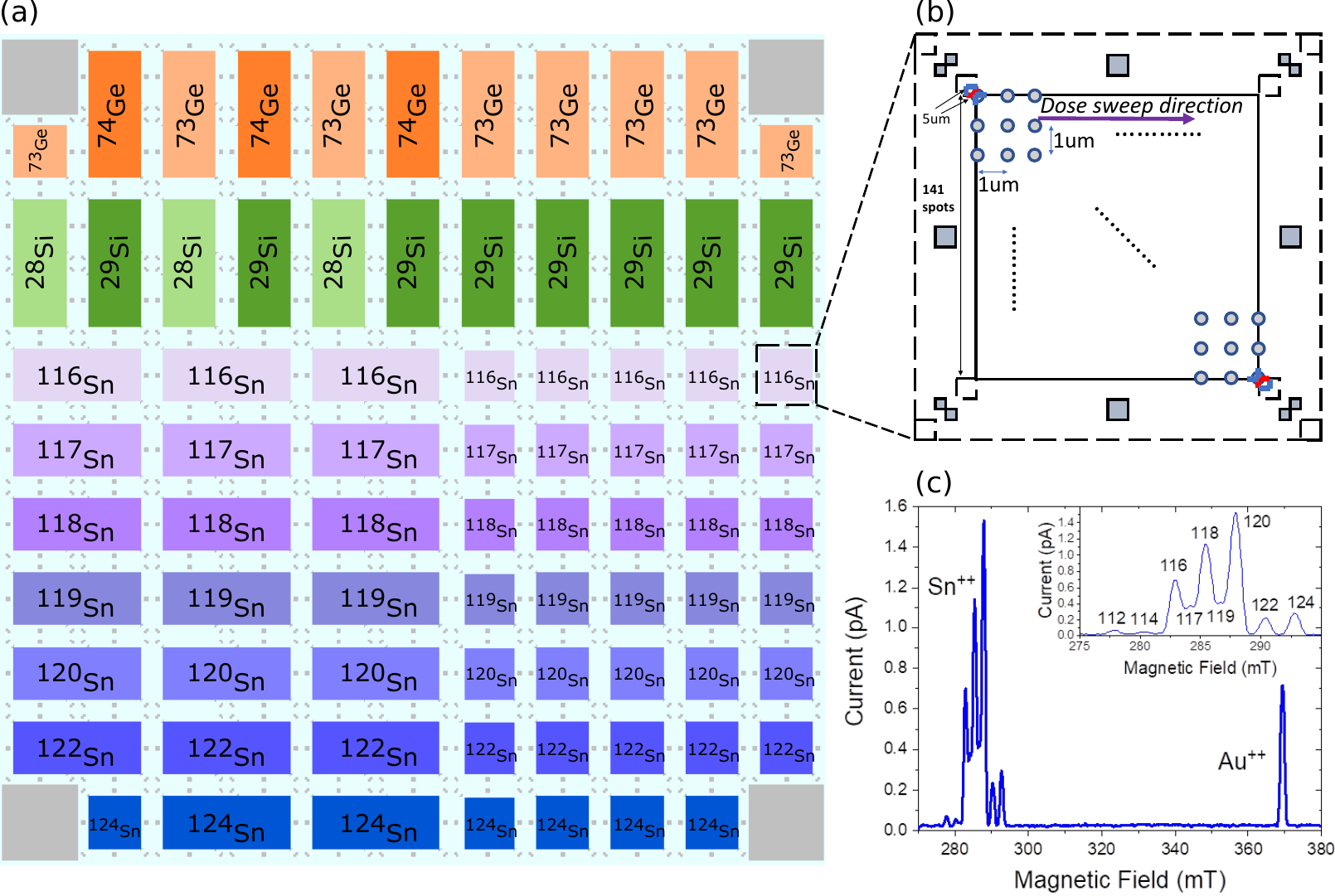}
    \caption{
    (a) Diagram of the isotope implantation of the UC1 sample, QR codes are shown as small gray squares within the implant regions.
    (b) Diagram of the grid implant pattern.
    (c) Mass spectrogram of the Sn isotopes.
    }
    \label{fig:sample}
\end{figure*}

The samples are implanted with group IV elements using a focused ion beam (FIB) tool at the Ion Beam Laboratory at Sandia National Laboratory.
Specifically, we selectively implanted different isotopes of Si, Ge, and Sn: \textsuperscript{28}Si, \textsuperscript{29}Si, \textsuperscript{73}Ge, \textsuperscript{74}Ge, \textsuperscript{116}Sn, \textsuperscript{117}Sn, \textsuperscript{118}Sn, \textsuperscript{119}Sn, \textsuperscript{120}Sn, \textsuperscript{122}Sn, and \textsuperscript{124}Sn in different regions aligned to the QR codes in the sample, as shown in Fig.~\ref{fig:sample}(a) for UC1. UC2 was implanted in a similar manner, with the arrangement of the implant zones varied to fit the slightly larger diamond plate.
Each element is implanted at two different energies corresponding to target implant depths of 25~nm and 75~nm.
The isotopic separation is not perfect, particularly for the Sn isotopes, which have a small relative isotopic mass difference, as shown in Fig.~\ref{fig:sample}(c).
We use the FIB's position selectivity to implant the dopants in a regular grid with 1~$\mu$m pitch, and a logarithmically increasing implant dose moving to the right in the sample, as shown in Fig.~\ref{fig:sample}(b).
The implant dose range is also swept from region to region, with doses swept between $10$ and $10^5$, $10^6$, $10^7$, \& $10^8$ ions per spot in different regions.
Hence, we have a mix of spin neutral and spin-ful isotopes for each group IV element in clearly marked, spatially separated regions.

Following the implantation, the dopants were converted to group-IV color centers via a high-temperature anneal at $1200^\circ$C in ultra-high vacuum ($<10^{-8}$~mBar) for the sample UC1, and a high-pressure high-temperature treatment ($>7~$GPa at 1950$^{\circ}$C) by Element Six.
This HPHT anneal is used to improve the emitter characteristics, narrowing the inhomogeneous distribution and improve charge stability~\cite{Gorlitz2020SpectroscopicDiamond,Narita2023}


\section{Experimental Setup}\label{sec:ple_methods}
The non-resonant photoluminescence (PL) measurements presented in Fig.~\ref{fig:Intro}(b-d) and the resonant, magnetic field-dependent measurements presented in Fig.~\ref{fig:GeV}(d) were taken on sample UC1, using setup one.
The wide-field photoluminescence excitation (WFPLE) experiments presented in Fig.~\ref{fig:GeV}(b,c) and Fig.~\ref{fig:SnV}(b-d) were taken on sample UC2 in setup two.
The magnetic field-dependent PLE measurements presented in Fig.~\ref{fig:SnV}(e) were taken on sample UC3, using setup three.
\subsection{Setup One}
\begin{figure*}
    \centering
    \includegraphics[width=\textwidth]{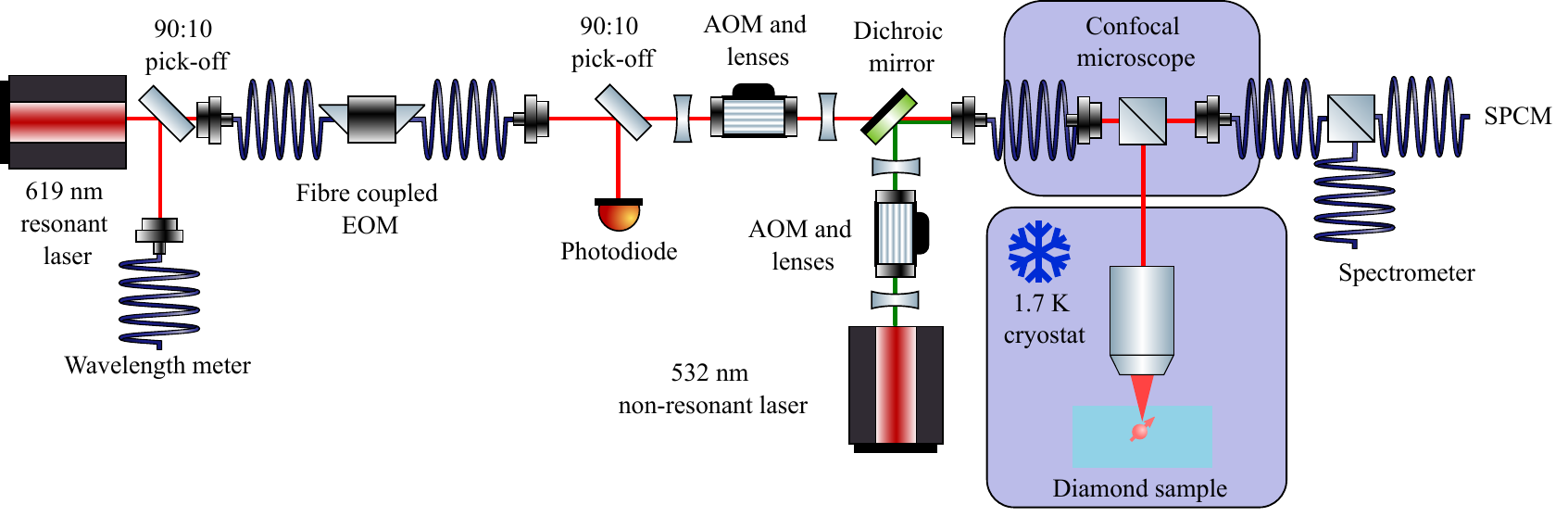}
    \caption{Diagram of setup one used for experiments on sample UC1.
    }
    \label{fig:setupone}
\end{figure*}

Setup one is outlined in Fig.~\ref{fig:setupone}. The diamond sample is housed in a closed-cycle cryostat (attoDRY
2100XL) with a base temperature of 1.7 K. Superconducting coils around the sample space facilitate the application of a two-axis vector magnetic field, with a maximum applied field of 9~T in the z direction, parallel to the length of the cryostat, and a maximum vector field of 3~T in the x direction, parallel to the base of the cryostat. The sample is situated on a three-axis piezoelectric stack  (ANPx101/LT and ANPz101/LT) used to move the area of interest beneath a 0.82 numerical aperture objective. A 4f lens system is installed within the cryostat and a confocal microscope sits atop the cryostat with a motorized mirror used to scan the excitation and collection paths $\approx$50~$\upmu$m$^2$ across the diamond sample to collect non-resonant hyperspectral maps. Non-resonant light is provided by a 532~nm laser pen (Roithner Lasertechnik CW532-100) whilst resonant light at 602~nm was provided by a Toptica DL SHG Pro, tunable between 600 and 620~nm. Resonant excitation was filtered from the collection path using a long-pass filter (Semrock BLP01-633R-25), tuned such that the cut-on frequency was 603~nm to collect only the phonon sideband (PSB) emission.

All resonant measurements in setup one were taken using chirped optical resonance excitation (CORE). A linearly chirped microwave pulse from an arbitrary waveform generator (Tektronix AWG70002A) (AWG) via a microwave amplifier
(Mini-Circuits ZX60-83LN12+) and a bias-tee (Mini-Circuits ZFBT-6G+) was applied to an electro-optic modulator (Jenoptik AM635), (EOM) to produce laser sidebands whose frequency was swept over several GHz within a few $\upmu$s. The EOM is kept locked to its interferometric minimum using a lock-in amplifier and PID loop (Red Pitaya, STEMlab 125-10). The collected PSB is sent to a single photon counting module (SPCM-AQRH-TR), which generates TTL pulses sent to a time-to-digital converter (Swabian Timetagger20). The start of a frequency sweep was triggered from the internal trigger of the AWG and also sent to the Timetagger20 to synchronize the measurement, creating a histogram where the time axis also represented the frequency sweep, matching the RF chirp. In this way, the resonance of an emitter could be swept over rapidly, avoiding any pumping effects and collecting statistics which could be used to remove drifts and emitter instability from the measurement. 

\subsection{Setup Two}
\begin{figure}
    \centering
    \includegraphics[width=0.8\columnwidth]{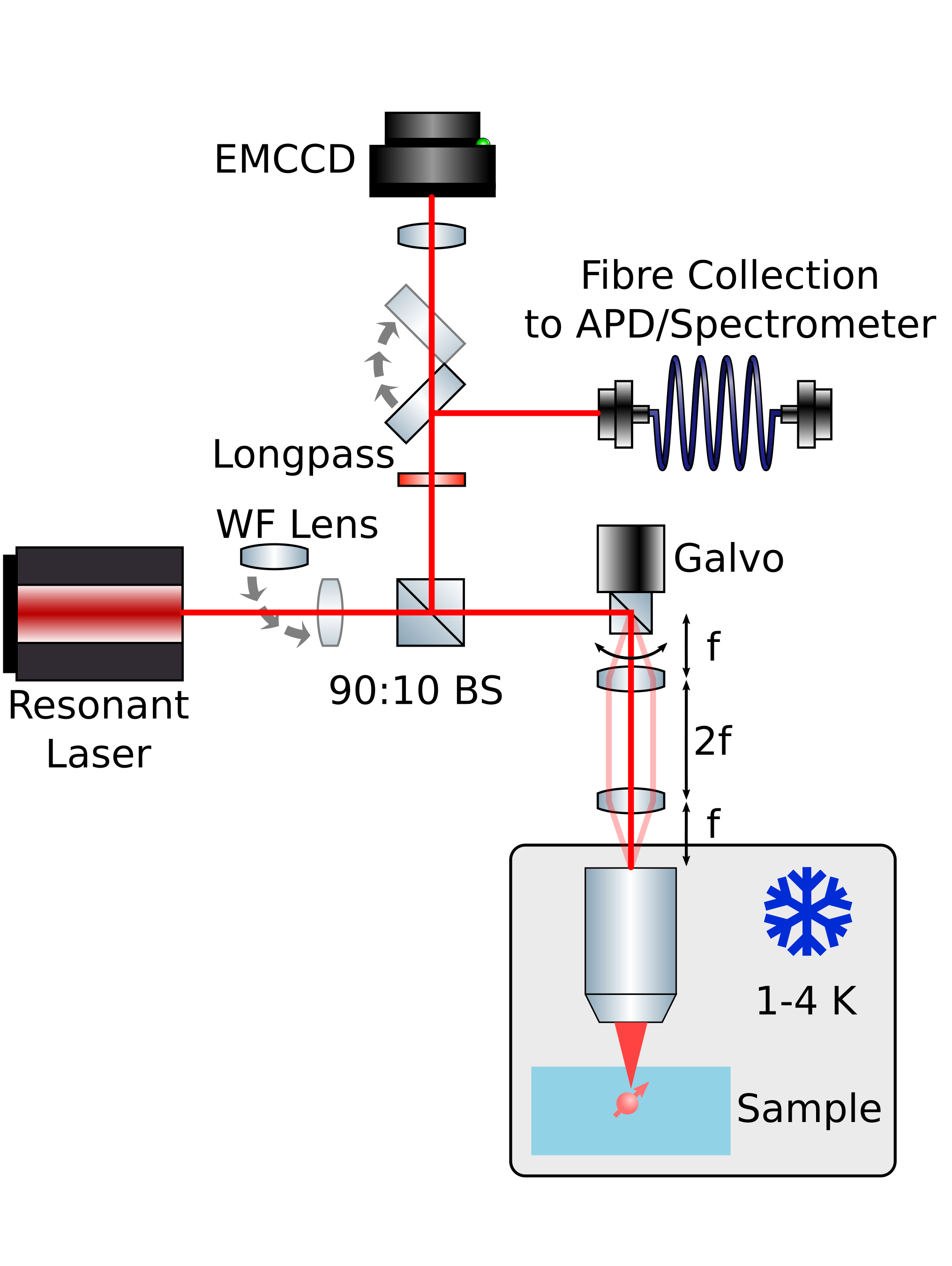}
    \caption{
    Diagram of the optical setup used for wide-field photoluminescence excitation experiments on sample UC2.
    }
    \label{fig:setup}
\end{figure}

WFPLE experiments were performed in a Montana systems cryostat at 5 K using the sample UC2.
A similar confocal setup was used, with the addition of a lens focused on the back plane of the 4f system to give widefield illumination, and a Photometrics Cascade 1K electron-multiplying charge-coupled device (EMCCD) camera to image the entire objective field of view. (see Fig.~\ref{fig:setup}).
WFPLE scans were performed at each selected implant region by scanning the M\textsuperscript{2} system output frequency across an approximately 16 GHz range in 4 MHz steps, taking an image at each step.
Location and frequency of all emitters in a field of view were labeled for each region, and the PLE spectrum generated from the WFPLE data by performing a Gaussian spatial average of the pixels surrounding the emitter location.
Fits were performed as described in the main text on this PLE data to extract emitter parameters.

\subsection{Setup Three}\label{sec: dil fridge}
Sample UC3 is a nanofabricated diamond sample implanted with isotopically-selected \textsuperscript{117}Sn~\cite{Parker2023}. The resonant measurements used the CORE technique through confocal collection and excitation, but the device was instead housed in a Bluefors LD250 He
dilution refrigerator at 3.5 K.

\section{Density Functional Theory Calculations}\label{sec:dft_calculations}
As discussed in the main text, the hyperfine interaction is parameterized by two matrices corresponding to the Fermi contact interaction ($\mathbf{A}_\mathrm{FC}$) and the dipole-dipole interaction ($\mathbf{A}_\mathrm{DD}$).
The Fermi contact term, $A_\mathrm{FC}$, depends only on the spin density at the nucleus, and so must be completely isotropic.
It can therefore be expressed as the identity matrix times the parameter $A_\mathrm{FC}$:
\begin{equation}\label{eq:fermicontact}
    A_\mathrm{FC} = \frac{2\pi}{3}g_\mathrm{I}\mu_\mathrm{N}g_\mathrm{e}\mu_\mathrm{B}\hbar^2\rho_\mathrm{S}(0)
\end{equation}
where $\rho_\mathrm{S}=\rho_\uparrow-\rho_\downarrow$ is the spin density, the difference between the density of spin up and spin down electrons $\rho_{\ud}$ around the defect.
The term $A_\mathrm{DD}$ depends on the spin-density distribution around the nucleus, and is therefore not isotropic in general.
It can be expressed as
\begin{equation}\label{eq:dipoledipole}
    \mathbf{A}_\mathrm{DD} = \frac{\mu_0}{4\pi}g_\mathrm{I}\mu_\mathrm{N}g_\mathrm{e}\mu_\mathrm{B}\hbar^2 \int d\mathbf{r} \left( \frac{\mathbf{r}\mathbf{r}}{r^5}-\frac{\mathbf{I}}{r^3} \right)\rho_\mathrm{S}(\mathbf{r})
\end{equation}
where $\mathbf{r}$ is the distance from the nucleus.

We use the QE-GIPAW package~\cite{QE2} within the Quantum Espresso plane-wave DFT code~\cite{QE1}.
Precise evaluation of Fermi contact and dipole-dipole interaction requires very accurate evaluation of the spin density in the vicinity of the nucleus.
The QE-GIPAW packages uses of the gauge-including projector augmented wave (GIPAW)~\cite{Pickard2001} method to reconstruct the valence electron orbitals near the nucleus, as well as core relaxation~\cite{Bahramy2007} to find the spin density at the nucleus.
We perform these calculations using the Perdew–Burke–Ernzerhof (PBE) functional~\cite{Perdew1996} with the defect in its \DDDD-symmetric configuration embedded in a 4x4x4 cubic (512 atom) supercell.

We also perform excited-state calculations to find the spin density and hyperfine properties for the $E_\mathrm{u}$ state.
These are performed with the same functional and supercell as the $E_\mathrm{g}$ calculation, however we use the $\Delta$SCF method to constrain the hole to one of the $e_\mathrm{u}$ Kohn-Sham orbitals~\cite{Thiering2018a, Ciccarino2020}.
To converge to the proper $E_\mathrm{u}$ state with the occupation constraint, we use a modified version of Quantum Espresso which implements the maximum overlap method~\cite{Gilbert2008} during the self-consistent field loop to update the occupations.

The PBE functional is generally accepted to not be as accurate as more sophisticated hybrid functionals such as HSE06, which are typically used in state-of-the-art DFT calculations of defect centers in diamond~\cite{Thiering2018a,Thiering2018b,Ciccarino2020,Harris2020a}.
PBE is known in particular to underestimate excited state energies compared to hybrid functionals.
However, as we are not concerned with the excited state energy and only with the distribution of spin near the nuclei in the ground and excited states, we find that the computationally cheaper PBE functional is sufficient, as evidenced by the reasonable agreement between the theory and experiment.
While switching to a hybrid functional may improve the accuracy, other effects such as finite supercell size, exclusion of the Jahn-Teller effect, and numerical errors in the projection of the PAW orbitals are likely larger contributors to the error in the hyperfine interaction calculations.

\section{Additional Electro-Nuclear Hamiltonian Terms}\label{sec:hamiltonian_details}
The total hyperfine Hamiltinonian is written as
\begin{equation}\label{eq:all_HF}
    \hat{H}_\mathrm{HF} = \hat{H}_\mathrm{FC} + \hat{H}_\mathrm{DD} + \hat{H}_\mathrm{IOC}  + \hat{H}_\mathrm{Q}
\end{equation}
where $\hat{H}_{FC}$ is the Fermi contact term, and $\hat{H}_\mathrm{DD}$ is the dipole-dipole term discussed in the main text.
We briefly discuss the additional terms $\hat{H}_\mathrm{IOC}$ and $\hat{H}_\mathrm{Q}$ shown in Eq.~\ref{eq:all_HF}.

The nuclear spin-orbit coupling (IOC) term, $\hat{H}_\mathrm{IOC}$, arises from the interaction of the nuclear spin with the magnetic field caused by the orbital motion of the hole around the defect center.
It can be expressed as
\begin{equation}\label{eq:IOC}
    \hat{H}_\mathrm{IOC} = - \frac{\mu_0}{4\pi}\frac{2g_\mathrm{I}\mu_\mathrm{B}\mu_\mathrm{N}}{\hbar^2} \frac{\mathbf{\hat{L}}\cdot\mathbf{\hat{I}}}{r^3}
\end{equation}
By analogy with the electronic spin-orbit coupling term, which is of the form $\hat{H}_{\mathrm{SOC}}=\lambda\mathbf{\hat{L}}\cdot\mathbf{\hat{S}}$~\cite{Hepp2014, Hepp2014a}, we can infer that in the basis defined in Eq.~\ref{eq:basis}, this reduces to
\begin{equation}
    \hat{H}_\mathrm{IOC} = \frac{1}{2}\upsilon \sigma_z^\mathrm{orb} \hat{I}_z
\end{equation}
where $\upsilon$ is a parameter quantifying the nuclear spin-orbit interaction.
This term results in an energy shift depending on the degree of alignment of the nuclear and orbital degrees of freedom.
At zero strain, this would change the spacing between the hyperfine levels to $\frac{1}{2}(A_\mathrm{FC}+A_\mathrm{DD}\pm\upsilon)$, with the sign depending on whether the hole is in the upper or lower branch.
First-principles calculations can be employed to find the numerical value of this parameter, however we expect it to be negligibly small for two reasons.
Firstly, the $r^{-3}$ component of Eq.~\ref{eq:IOC} greatly suppresses the IOC interaction to be at most roughly as strong as the dipole-dipole interaction ($<5\%$ of the total hyperfine interaction).
Secondly, the orbital magnetic term in Eq.~\ref{eq:hamiltonian}, $\hat{H}_\mathrm{L} = q\mu_\mathrm{B} \mathbf{\hat{L}} \cdot \mathbf{B}$, is known to have a small effective response $q\approx 0.1$.
This can be attributed to a decreased effective orbital angular momentum due to the presence of the lattice.
We would expect this effect to further decrease the IOC, making this term roughly 3 orders of magnitude smaller than the total hyperfine interaction discussed in the paper.
We leave it to future theoretical and experimental work to identify the value of this nuclear spin-orbit parameter $\upsilon$.

The quadrupole interaction term $\hat{H}_\mathrm{Q}$ arises from the non-spherical charge distribution of nuclear spins with $I>1/2$, and is zero for $I\leq 1/2$.
It is therefore only applicable to the spin-9/2 isotope \textsuperscript{73}Ge discussed in this paper.
The quadrupole interaction term results in a energy shift dependent on the gradient of the electric field at the location of the nucleus, and takes the form
\begin{equation}
    \hat{H}_\mathrm{Q} = \mathbf{\hat{I}} \cdot \mathbf{Q} \cdot \mathbf{\hat{I}}
\end{equation}
where $\mathbf{Q}$ is a matrix proportional to the electric field curvature at the nucleus, and the nuclear quadrupole moment.
If no external field is applied, the electric field curvature can only come from the electronic distribution around the defect.
Similar to the hyperfine dipole-dipole interaction matrix $\mathbf{A}_\mathrm{DD}$, the defect's \DDDD\ symmetry restricts the matrix $\mathbf{Q}$ to be diagonal in the basis defined in Eq.~\ref{eq:basis}, with $-2Q_{xx}=-2Q_{yy}=Q_{zz}=Q$.
In the large electronic spin-orbit coupling limit discussed in Section~\ref{sec:DFT}, the quadrupole coupling creates an anharmonic shift of $Qm_\mathrm{J}^2$ of the eigenstates $\ket{\ud}\ket{m_\mathrm{J}}$.
If the nuclear spin sublevels are to be addressed individually, the anharmonicity sets the speed at which they can be manipulated, as a Rabi frequency exceeding the anharmonicity will cause unwanted driving of the wrong transition.
From DFT, we estimate the parameter $Q=4.3$ MHz for \textsuperscript{73}GeV\m\ in the ground state, and therefore only causes a shift of the energy levels which is much smaller than the large $A_\mathrm{FC}$ Fermi contact.
We are therefore justified in neglecting it in the main discussion in the paper.

Finally, we briefly discuss the effect of the Jahn-Teller distortion on the defect.
The Jahn-Teller effect is a symmetry-breaking distortion that occurs due to unequal occupation in degenerate orbitals.
In the case of the group-IV color center defects, the Jahn-Teller effect causes a distortion from \DDDD\ to one of three equivalent \CHH-symmetric configurations.
This distortion affects the distribution of the spin density, $\rho_S$, around the dopant nucleus.
The hyperfine interaction parameters $A_\mathrm{FC}$ in Eq.~\ref{eq:fermicontact} and $A_\mathrm{DD}$ in Eq.~\ref{eq:dipoledipole} are therefore affected by this distortion, resulting in a shift of the isotropic term:
\begin{equation}
    A_\mathrm{FC}^{C_{2h}}=A_\mathrm{FC}^{D_{3d}} + \delta A_\mathrm{FC}
\end{equation}
and a shift of the anisotropic terms
\begin{equation}
    \begin{split}
        A_{xx/yy}^{C_{2h}} & = -\frac{1}{2}\left( A^{D_{3d}}_{DD} + \delta A_\mathrm{DD} \right) \pm \delta A_\perp \\
        A_{zz}^{C_{2h}}    & = A^{D_{3d}}_{DD} + \delta A_\mathrm{DD}
    \end{split}
\end{equation}

We estimate the size of these parameters by comparing the hyperfine parameters for SiV\m\ in the \DDDD\ and \CHH\ configuration in a 3x3x3 cubic (216 atom) supercell, with the results shown in Table \ref{tab:JT}
\begin{figure*}
    \centering
    \includegraphics[width=\textwidth]{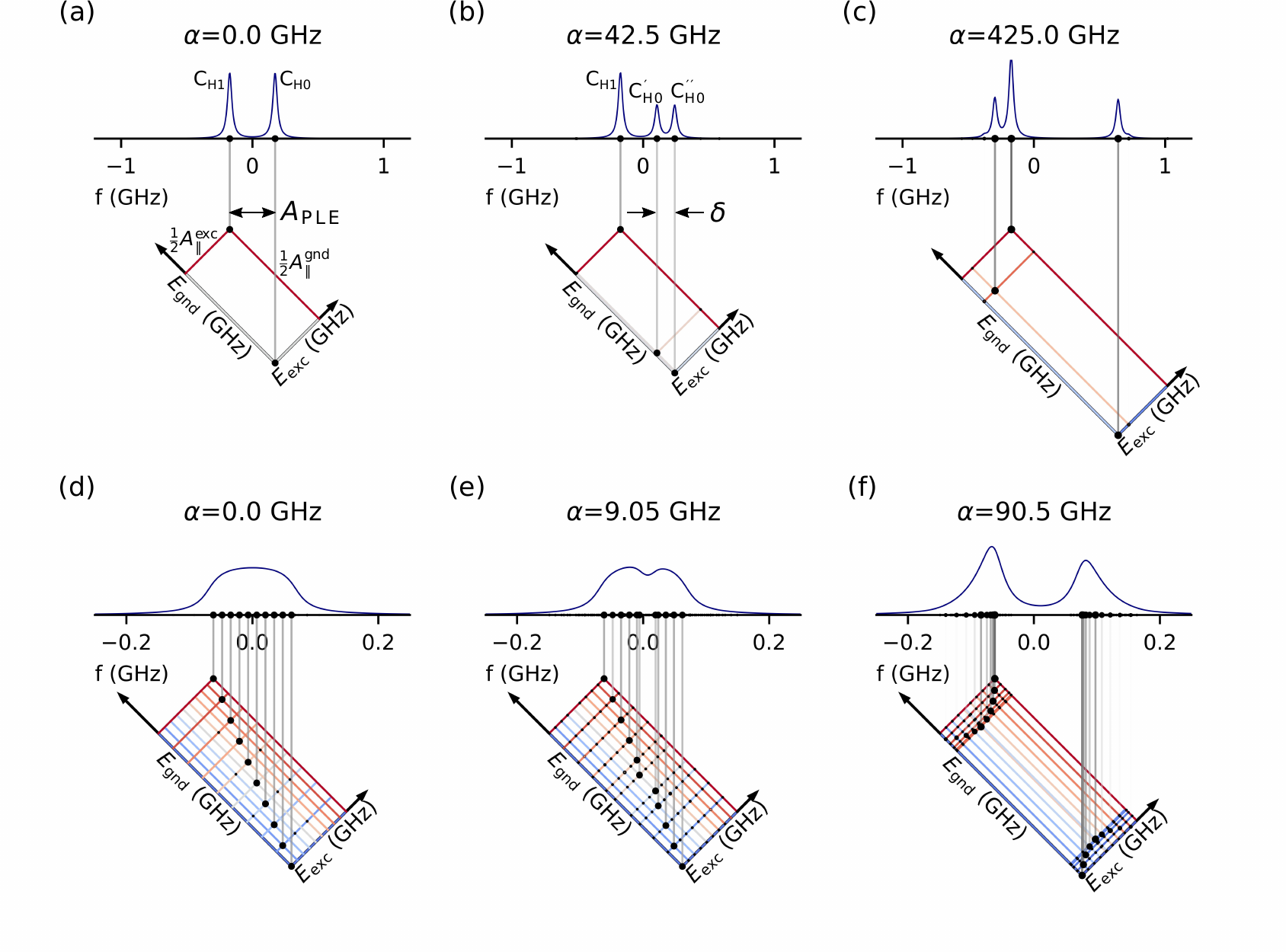}
    \caption{
    Visualization of spectra for (a, b, c) \textsuperscript{117}SnV\m\, and (c, d, e) \textsuperscript{73}GeV\m\ using predicted hyperfine parameters at various strains, $\alpha$.
    The bottom half of each plot shows the hyperfine levels in the ground/excited state.
    At each intersection a line whose intensity is proportional to the strength of the corresponding transition is plotted.
    Each transition line aligns with the relative frequency of the optical transition, and the resulting predicted spectrum is shown in the upper half of the plot.
    }
    \label{fig:transitions}
\end{figure*}
\begin{table}
    \centering
    \begin{tabular}{c|c|c|c|c}
        Config. & $A_\mathrm{FC}$ (MHz) & $A_\mathrm{DD}^{xx}$ (MHz) & $A_\mathrm{DD}^{yy}$ (MHz) & $A_\mathrm{DD}^{zz}$ (MHz) \\ \hline
        \DDDD & 54.4 & 1.17 & 1.17 & -2.34 \\
        \CHH & 49.6 & -0.456 & -0.456 & 3.16
    \end{tabular}

    \caption{Comparison of the hyperfine coupling for \textsuperscript{29}SiV\m\ in the \DDDD\ and \CHH\ configurations}
    \label{tab:JT}
\end{table}
For the 216 atom supercell, the results imply
$\delta A_\mathrm{FC}=-4.8$~MHz, $\delta A_\mathrm{DD} = 5.5$~MHz, and $\delta A_\perp < 0.01$~MHz.
The distortion manifests mostly as a shift in the \DDDD\ values, with the symmetry-breaking component $\delta A_\perp$ remaining very small, giving some insight into why the $C_\mathrm{H0}$ line does not split at zero strain due to Jahn-Teller.
Given these results, we estimate that Jahn-Teller distortion changes the magnitude of the hyperfine coupling by approximately 10\% of the total value.
This potentially explains a portion of the error in the DFT calculations in the main text, where Jahn-Teller is neglected.
A complete treatment of the Jahn-Teller effect would require inclusion of the Jahn-Teller phonon modes to find the combined spin-phonon vibronic states~\cite{Thiering2018a}, where the effective hyperfine interaction would then be averaged over the vibronic configurations.
Since Jahn-Teller does not qualitatively change the predictions, we leave a detailed study for future work.

\section{Visualization of Hyperfine Spectra}\label{sec:PLE_vis}

In Fig.~\ref{fig:transitions} below, we show a plot highlighting the origin of the hyperfine optical spectrum.
\begin{figure}
    \centering
    \includegraphics[width=\columnwidth]{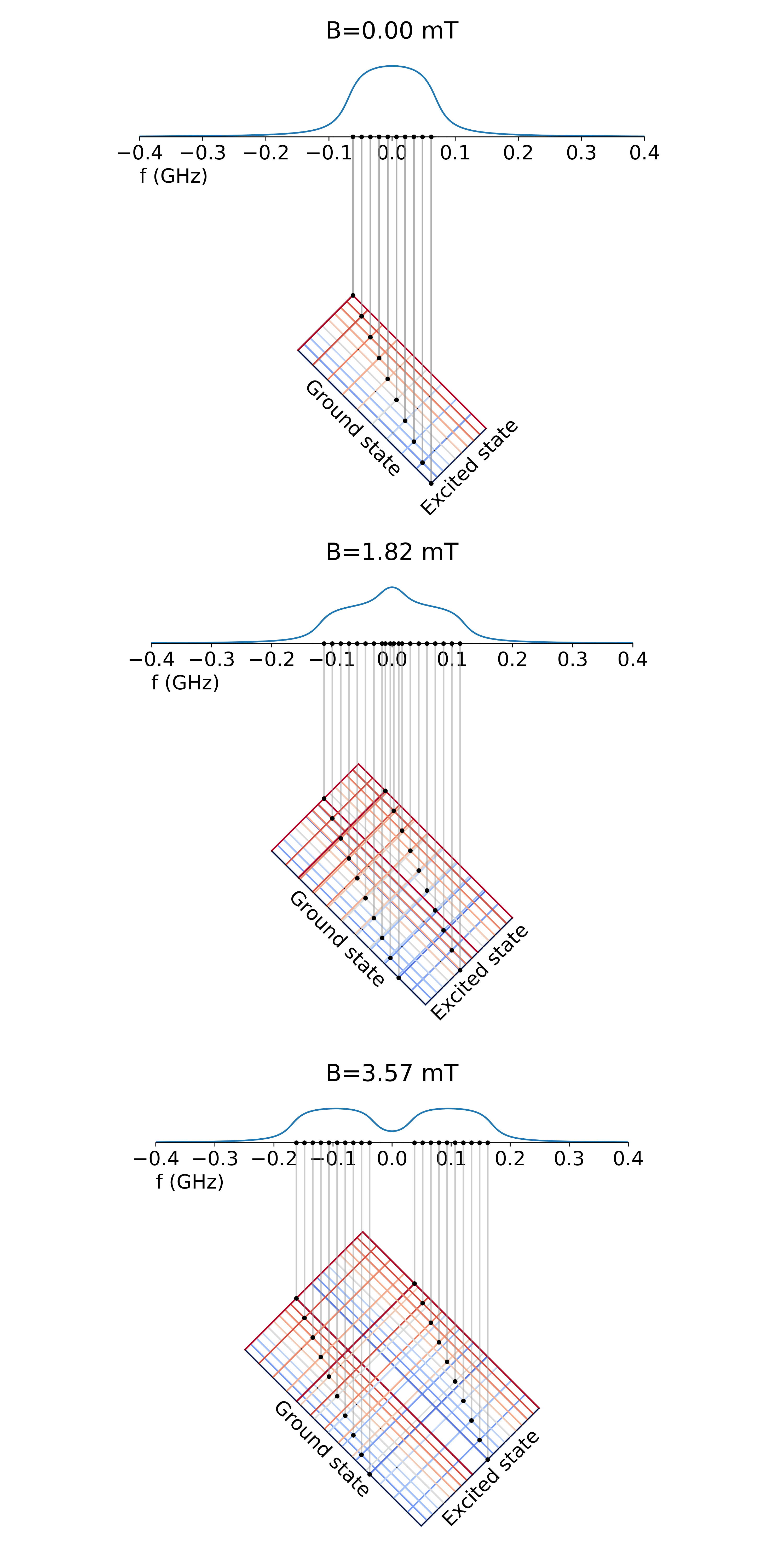}
    \caption{
    Visualization of spectra for \textsuperscript{73}GeV\m\ as a function of B-field.
    }
    \label{fig:transitions2}
\end{figure}
In the lower half of each subplot, we plot the ground state and excited state hyperfine energy levels as a set of lines, with color corresponding to the total angular momentum $\langle J^2\rangle$ of that state (red being nuclear and electron spins aligned, blue anti-aligned).
The intersection of each of the ground/excited state lines corresponds to a potential transition which can occur, and a line is plotted for each transition to the upper axis, with the intensity of the line corresponding to the predicted intensity of that transition.
The lower half plot is rotated and scaled in such a way that the position along the x-axis of the upper half plot lines up with the transition frequency.
A predicted spectrum is shown, calculated as the intensity-weighted sum of all the transitions.

It is evident that the transitions with the largest contribution are the ones between states with identical $\langle J^2\rangle$, i.e. spin-conserving transitions.
As the strain increases, the ground state hyperfine levels start to split as discussed in Section~\ref{sec:DFT}.
This effect is suppressed in the excited state due to the higher spin-orbit coupling in the $E_\mathrm{u}$ state.

For \textsuperscript{117}SnV\m\ in Fig.~\ref{fig:transitions}(a-c), the spectrum at zero strain consists of two peaks corresponding to transitions between the ground and excited $m_\mathrm{J}=\pm1$, and $m_\mathrm{J}=0$ levels respectively.
At higher strain the $m_\mathrm{J}=0$ peaks split apart primarily due to the ground level's strain splitting, which can be seen in the spectrum in Fig.~\ref{fig:SnV}(c).
The spectra in Fig.~\ref{fig:SnV}(b) are averaged over many sites, with the frequency shifted such that the lowest energy peak is at 0.
The higher frequency peak is then reduced in intensity due to the combined effects of: (1) residual spin-0 isotopes which don't have the secondary peak at all, and (2) variability in the hyperfine optical splitting parameters $A_\mathrm{PLE}$ and $\delta$ from measurement noise and differences in strain between emitters.

For \textsuperscript{73}GeV\m\ in Fig.~\ref{fig:transitions}(d-f), the strain causes the flat-topped peak to split into two peaks corresponding to the transitions between the $J=5$ levels at lower energy, and $J=4$ levels at higher energy.
We note that the predicted spectrum is shown for a near lifetime-limited linewidth, and these features are not easily visible in the WFPLE GeV\m\ spectra shown in the main text.

We also highlight the effect of a magnetic field aligned along the z-direction on \textsuperscript{73}GeV\m\ in Fig.~\ref{fig:transitions2}.
As the B-field increases the states containing terms with electronic states $\ket{\uparrow}$ and $\ket{\downarrow}$ are shifted in opposite directions.
The ground and excited state are shifted by differing amounts due to the difference in orbital magnetic field response in the ground (excited state), $q_{gnd(exc)}$.
Note that $q_{exc}$ has been artificially decreased to keep the figure compact.
The same general phenomenon occurs with the true experimental value, albeit at the higher fields observed in Fig.~\ref{fig:GeV}(d).

As the magnetic field increases, the two groups of transitions stop overlapping completely as they do at zero field.
For sufficiently strong fields, the two groups of transitions become fully separated, and two flat-topped peaks become apparent in the spectrum.
At intermediate field strengths, the transitions near zero detuning still overlap, causing a characteristic bump at the center of the transition peak, with two broad shoulders.

\end{document}